\documentclass{mn2e}
\usepackage{graphicx}

\def \sanch{S\'anchez-Bl\'azquez}

\def \kms {${\rm{km}\,\rm{s}^{-1}}$}

\def \ha   {H\,$\alpha$}

\def \afe  {[$\alpha$/Fe]}

\def \zh  {[Z/H]}

\def \eza   {{\sc EZ-ages}}

\def \Hbeta{Hbeta}
\def \HgF{HgF}
\def \HdF{HdF}

\title[Recently-quenched dwarfs in the outskirts of Coma]
{A large population of recently-quenched red-sequence dwarf galaxies in the 
outskirts of the Coma Cluster\thanks{Observations reported here were obtained at the MMT Observatory, 
a joint facility of the University of Arizona and the Smithsonian Institution.}}
\author[Russell J. Smith et al. ]
{Russell J. Smith$^{1}$, 
Ronald O. Marzke$^{2}$, 
Ann E. Hornschemeier$^{3}$, 
Terry J. Bridges$^{4}$, 
\newauthor
Michael J. Hudson$^{5}$, 
Neal A. Miller$^{6}$, 
John R. Lucey$^{1}$ 
Gerardo A. V\'azquez$^{6,7}$,
\newauthor
David Carter$^{8}$
\\
$^1$Department of Physics, University of Durham, Durham DH1 3LE\\
$^2$Department of Physics and Astronomy, San Francisco State University, San Francisco, CA 94132, USA\\
$^3$NASA Goddard Space Flight Centre, Code 662.0, Greenbelt, MD 20771, USA\\
$^4$Department of Physics, Queen's University, Kingston, Ontario K7L 3N6, Canada\\
$^5$Department of Physics and Astronomy, University of Waterloo, 200 University Avenue West, Waterloo, Ontario N2L 3G1, Canada\\
$^6$Department of Physics and Astronomy, Johns Hopkins University, 3400 North Charles Street, Baltimore, MD 21218, USA\\
$^7$Department of Physics, Salisbury University, 1101 Camden Avenue, Salisbury, MD 21801, USA\\
$^8$Astrophysics Research Institute, Liverpool John Moores University, Twelve Quays House, Egerton Wharf, Birkenhead CH41 1LD
}

\date{Accepted 3rd March 2008}

\pagerange{\pageref{firstpage}$-$\pageref{lastpage}} \pubyear{2008}

\begin{document}

\label{firstpage}

\maketitle

\begin{abstract}
We analyse the stellar populations of 75 red-sequence 
dwarf galaxies in the Coma cluster, based on high signal-to-noise spectroscopy from the 
6.5m MMT. 
The sample covers a luminosity range 3--4 magnitudes below $M^\star$, in the cluster 
core and in a field centred 1\,deg to the south-west. 
We find a strong dependence of the absorption line strengths with location in the cluster. 
Galaxies further from the cluster centre have 
stronger Balmer lines than inner-field galaxies of the same luminosity. The magnesium lines
are weaker at large radius, while the iron lines are not correlated with radius. 
Converting the line strengths into estimates of stellar age, metallicity and abundance ratios, we find the 
gradients are driven by variations in age ($>6\sigma$ significance) and in the iron abundance Fe/H ($\sim2.7\sigma$ significance). 
The light element (Mg, C, N, Ca) abundances are almost independent of radius. 
At radius of 0.4--1.3\,degree ($\sim$0.3--1.0$\times$ the virial radius), dwarf galaxies have ages $\sim$3.8\,Gyr on average, 
compared to $\sim$6\,Gyr near the cluster centre. 
The outer dwarfs are also $\sim$50\% more iron-enriched, at given luminosity.
Our results confirm earlier indications that the ages of red-sequence galaxies depend on location within clusters, and
in Coma in particular. The exceptionally strong trends found here suggest that dwarf galaxies are especially
susceptible to environmental ``quenching'', and/or that the south-west part of Coma is a particularly clear example of 
recent quenching in an infalling subcluster. 
\end{abstract}

\section{Introduction}
Galaxies in rich clusters are subject to a range of environmental processes (stripping, suffocation, harrassment, tides) which are absent, 
or at least less efficient, in the field and in poor groups. The role of these effects in ``quenching''
star formation in cluster galaxies has been 
widely discussed based on instantaneous star-formation tracers such as \ha\ emission. For example, 
the fraction of star-forming galaxies is supressed out to $\sim$3 times the virial radius $R_{200}$ (e.g. Lewis et al. 2002). 
To trace the history of environment-driven quenching in today's galaxy clusters, high quality spectroscopy is needed 
to determine the final epoch of star-formation in now-passive galaxies. 

A number of works have claimed to observe significant variation in absorption line strengths of galaxies, 
as a function of radius in clusters (e.g. Guzm\'an et al. 1992; Carter et al. 2002; Smith et al. 2006). 
These trends have been interpreted as a radial gradient in average stellar age. 
Smith et al. note evidence that dwarfs may exhibit stronger radial dependences than the more massive galaxies, as expected
if the efficiency of the quenching process depends on the depth of the potential well binding the gas reservoirs. 

In this Letter, we provide further evidence for strong cluster-centric gradients in
age for dwarf galaxies (median $M_r\approx{}-17.5$) in Coma, using new 
spectroscopic data from the 6.5m MMT. 
A comprehensive analysis of the stellar populations will appear in a future paper (Smith et al., in preparation). 
For luminosity and distance calculations we adopt a distance to Coma of 100\,Mpc.

\section{Observations and parameter measurements}

\begin{figure}
\includegraphics[angle=270,width=170mm]{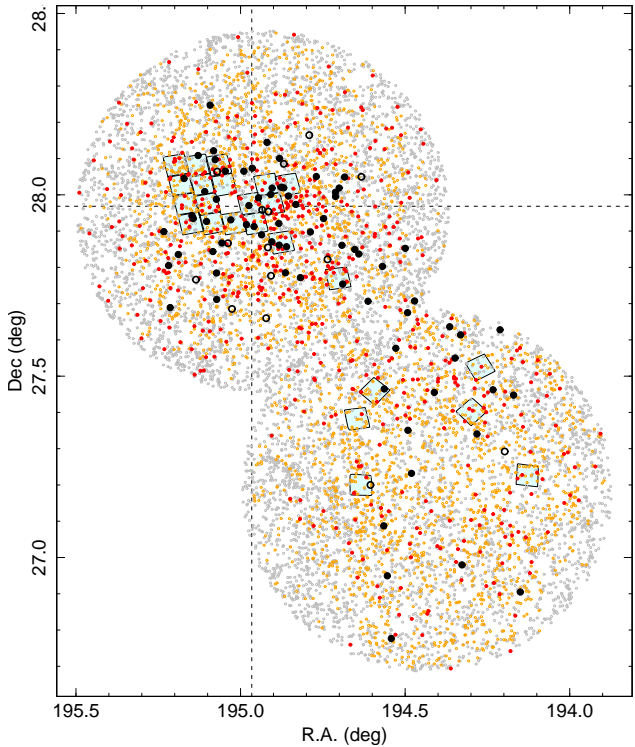}
\vskip -0mm
\caption{Sky distribution of the linestrength sample (black). The filled points are galaxies used in 
the fits of Section~\ref{sec:trends}. Open symbols are either bright ($M_r<-19$) comparison galaxies, have emission lines, or no measured age. 
Small points show other SDSS galaxies within the Hectospec survey region in red (confirmed members), yellow (confirmed 
background galaxies) or grey (no redshift information). 
The redshifts are from our Hectospec survey and literature sources. Pale blue
squares indicate the regions covered by HST imaging from the Coma ACS Treasury Survey (Carter et al. 2008).
Crosshairs show the adopted cluster centre, midway between the two central cD galaxies. 
}
\label{fig:samplesky}
\end{figure}

\begin{figure}
\includegraphics[angle=270,width=85mm]{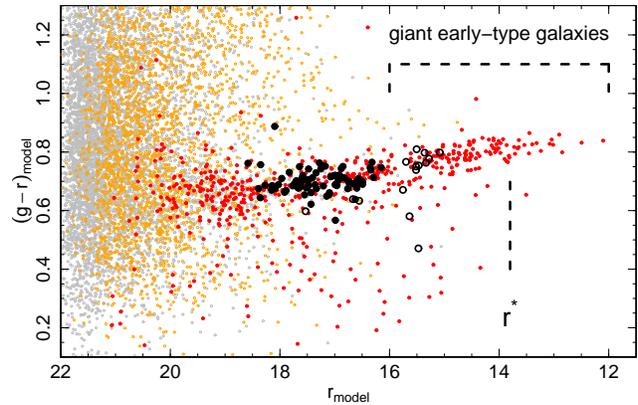}
\vskip -0mm
\caption{
The $g-r$ colour magnitude diagram for the galaxies in Figure~\ref{fig:samplesky}, from SDSS photometry. 
For comparison, we note the position of the LF break, $r^*$, and the magnitude range for ``giant'' 
early-type galaxies, having velocity dispersions $\sigma\ga75$\,\kms. Symbols are as in Figure~\ref{fig:samplesky}.}
\label{fig:samplecmr}
\end{figure}

\begin{figure*}
\includegraphics[angle=0,width=180mm]{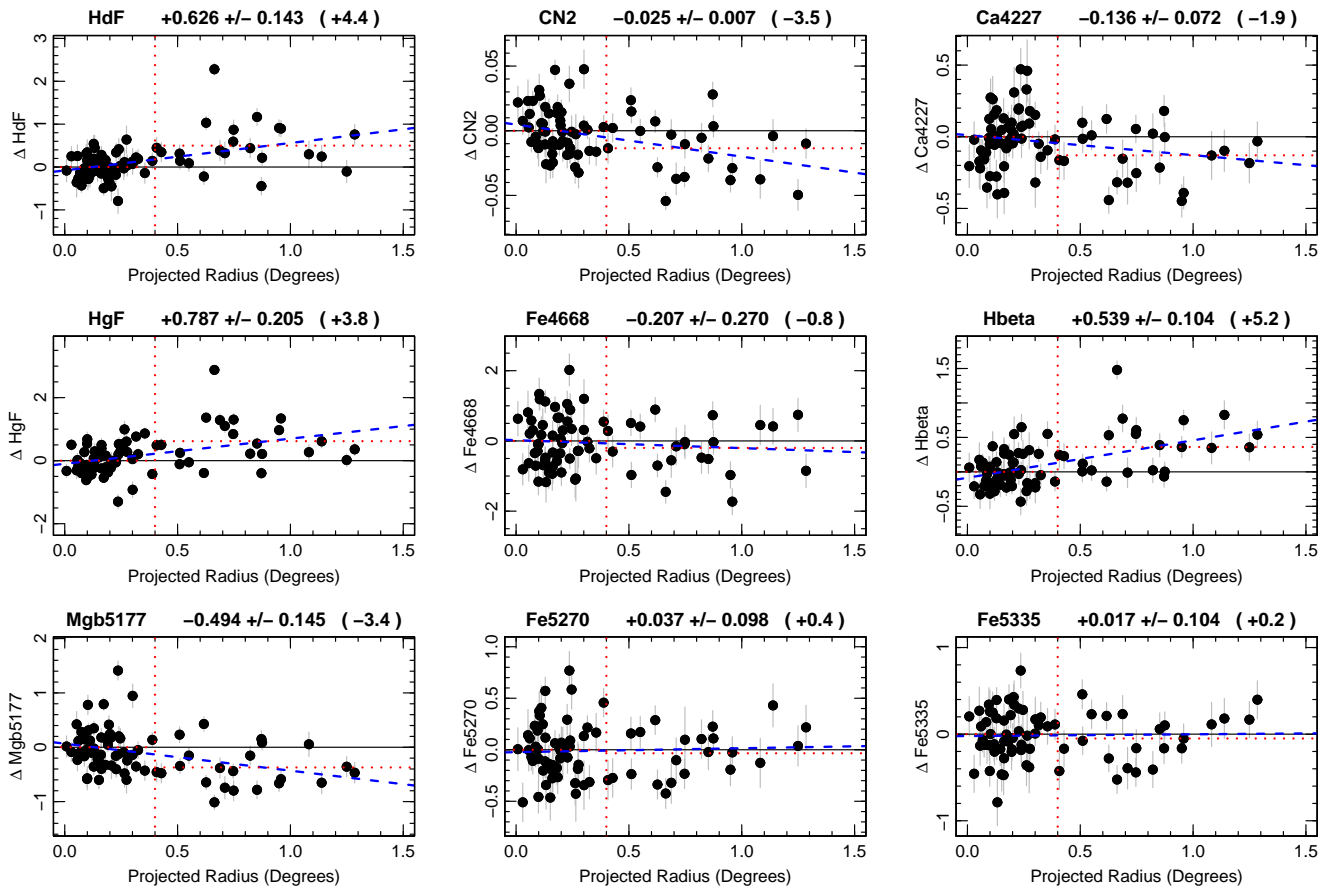}
\vskip -0mm
\caption{Radial trends in the index residuals. The title bar for each index $I$ gives the coefficient of 
projected radius $R_{\rm proj}$ in a bivariate fit of the form $I = a_0 + a_1 M_r + a_2 R_{\rm proj}$. 
The significance of the radial trend is given in parentheses in units of the standard error. The red dotted lines show median residuals inside
and outside 0.4\,deg, while the dashed blue line shows an unweighted fit to the residuals.}\label{fig:radtrends_ix}
\end{figure*}

\begin{figure*}
\includegraphics[angle=0,width=180mm]{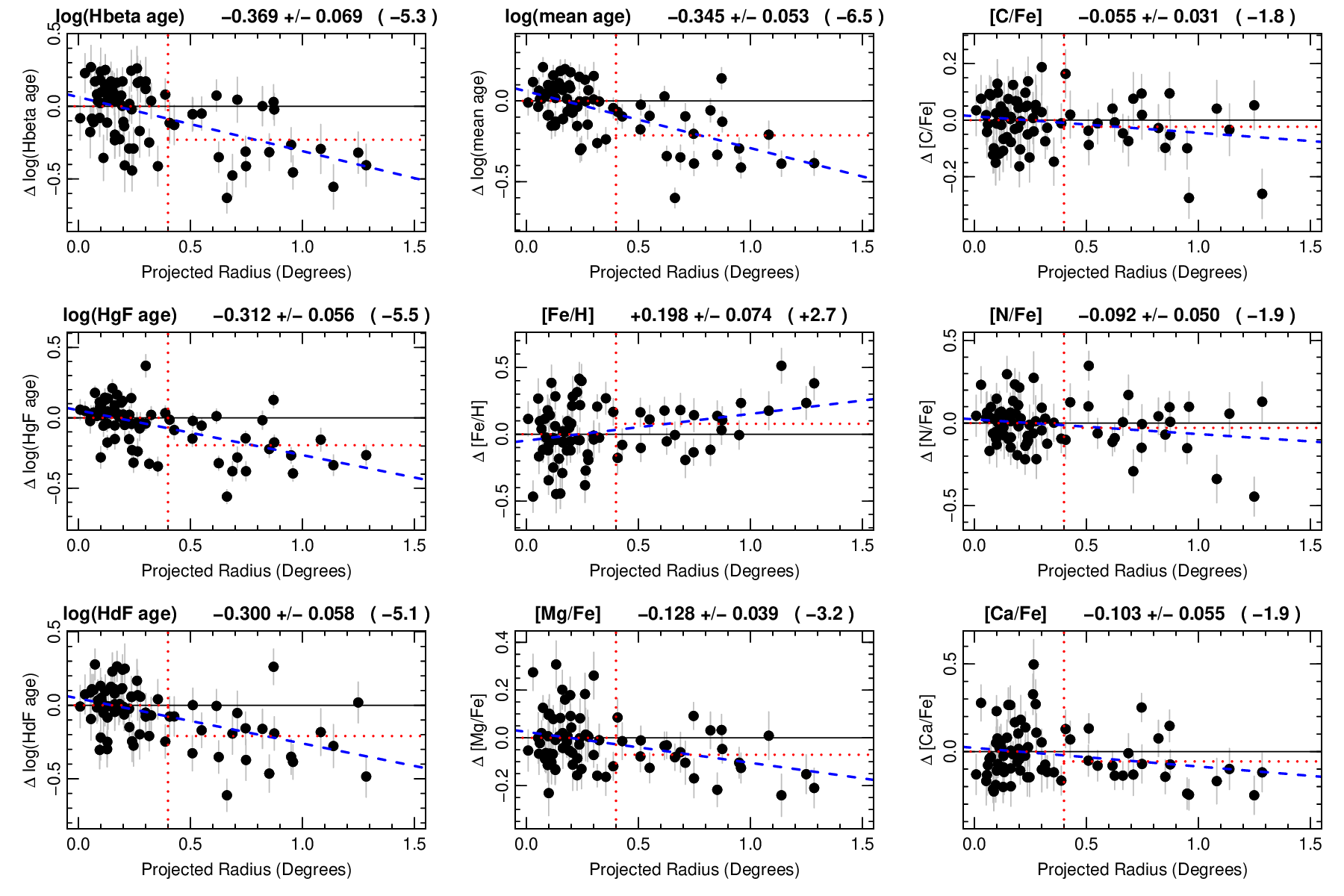}
\vskip -0mm
\caption{Radial trends in the stellar population parameters from \eza. Annotations as in Figure~\ref{fig:radtrends_ix}.}\label{fig:radtrends_ez}
\end{figure*}

High signal-to-noise spectra were obtained using the Hectospec fibre-fed spectrograph on the MMT, in February--April 2007.
The instrument deploys 300 fibres over a 1\,deg diameter field of view (corresponding to 1.75\,Mpc at Coma); 
the fibre diameter is 1.5\,arcsec (0.7\,kpc). 
Two fields were observed, one centred on the cluster core, and an outer field to the south-west (Figure~\ref{fig:samplesky}). 
The choice of the south-west field was motivated by available supporting data, and influenced by 
the work of Caldwell et al. (1993) who reported a high fraction of Balmer-strong galaxies in this region. 

Our observations were made in parallel with an extensive redshift survey of the cluster (Hornschemeier et al., in preparation). 
To study the stellar populations of dwarf galaxies, we observed 79 known cluster members with luminosities $3-4$\,mag below $M^*$, 
plus ten brighter galaxies for overlap with previous studies (e.g. \sanch\ et al. 2006). 
The target galaxies were selected to lie close to the red sequence of non-star-forming galaxies (Figure~\ref{fig:samplecmr}).
The 270\,line\,mm$^{-1}$ grating was used, resulting in a wide wavelength coverage (3700--9000\,\AA) at 
a  spectral resolution of 4.5\,\AA, FWHM. The median total integration time for the faint galaxies was $\sim$7\,hours, yielding
typical signal-to-noise ratio of $\sim$40\,\AA$^{-1}$ (at $\sim$5000\,\AA). 
The brighter galaxies were observed for 0.7--2.0 hours. 
Relative flux calibration was imposed using F stars with  photometry from Sloan Digital Sky Survey 
(SDSS, Adelman-McCarthy et al. 2007), 
observed simultaneously with the galaxies in each configuration. 
The data were reduced using {\sc hsred}, an automated {\sc idl} package based on the SDSS pipeline, 
provided by Richard Cool.

Analysis of the one-dimensional spectra, including combination of multiple exposures and measurements of emission and absorption lines, followed methods  
used by Smith, Lucey \& Hudson (2007). 
Three galaxies with \ha\ in emission (equivalent width $\ga$20\,\AA) were removed from the sample. 
No other galaxies have emission above $\sim$0.5\,\AA\ after removing the stellar continuum. 
Absorption-line indices were measured and corrected to the Lick resolution, but not transformed to the Lick ``system'', 
since we will use models based on flux-calibrated spectral libraries. 
Comparisons with the index measurements of \sanch\ et el. (2006) show excellent agreement
for eight galaxies in common 
(e.g. \HgF\ offset $0.1\pm0.1$\,\AA\ with rms scatter 0.2\,\AA). 

Line strength measurements can be transformed into estimates of stellar population age and element abundances, through 
comparison with population synthesis models. In this Letter, we analyse the absorption line measurements with reference to 
new stellar population models by Schiavon (2007). 
Advantages of this model set include its basis in a flux-calibrated stellar library, 
individually variable Mg, C, N and Ca abundances at fixed Fe/H, and a publicly-available 
code \eza\ for estimating parameters from measured indices (Graves \& Schiavon 2008). 

The \eza\ code performs an iterative ``sequential'' grid inversion, finding abundance ratios which 
yield consistent age and metallicity estimates across a range of index--index diagrams. 
A full description of the method is 
given by Graves \& Schiavon (2008), while a summary can be found in Schiavon (2007). 
In our implementation, an initial estimate of age and Fe/H is made using \Hbeta\ and the iron 
indices Fe5270 and Fe5335 assuming solar-scaled abundances. 
The Mg/Fe ratio is adjusted to yield the same metallicity and age 
from \Hbeta\ and Mgb5177. Similarly, C/Fe is obtained using the Fe4668 index (which is heavily
inflenced by the C abundance). With C/Fe in hand, the N/Fe abundance is adjusted for consistency 
with the measured CN2. Finally, Ca/Fe is obtained from Ca4227. The procedure is iterated, 
deriving a new age and metallicity estimate from the updated abundance pattern in each step. 
Once the final abundances are obtained, ages are estimated using the higher-order Balmer indices 
\HdF\ and \HgF, as well as using \Hbeta. Errors are determined from Monte Carlo simulations.

\section{Radial trends}\label{sec:trends}

In this section, we examine the evidence for trends in our sample of Coma dwarfs, both at the
level of observed indices and also using the derived ages and metallicities. Because the 
stellar populations of red galaxies follow strong scaling relations with the mass or luminosity 
(e.g. Smith et al. 2007), we analyse the residuals from the index--luminosity (and age--luminosity relations, etc), 
rather than the index data themselves. Thus we are comparing the inner and outer galaxy population 
at fixed luminosity. (In fact, there is no correlation of luminosity with radius within our
galaxy sample, so similar results would be obtained without this control.) 
We exclude the brighter comparison galaxies (with $r<16$), and one object for which \eza\ failed to converge, 
leaving 75 galaxies in the fits. 
All galaxies are assigned equal weight in the fits, 
since the scatter is primarily intrinsic and the measurement errors similar for all points. 

The residual index--radius relations are shown in Figure~\ref{fig:radtrends_ix}, for the nine indices which 
enter into our analysis with \eza. Strong positive correlations are observed in all of the Balmer indices: 
\Hbeta\ (5$\sigma$), \HgF\ (4$\sigma$), and \HdF\ (4$\sigma$). 
Negative, but generally less significant, trends are recovered for the metal indices dominated by light elements: 
CN2 (3$\sigma$), Mgb5177 (3$\sigma$), Ca4227 (2$\sigma$).
Finally, the carbon-dominated index Fe4668 and iron-dominated 
Fe5270 and Fe5335 show no significant correlation.

\begin{figure}
\vskip 3mm
\includegraphics[angle=270,width=85mm]{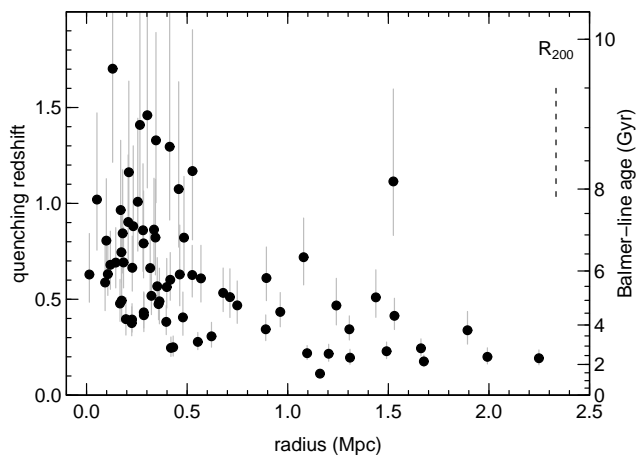}
\vskip -2mm
\caption{Quenching redshift, defined here as the redshift for which the look-back time is equal to the Balmer age,
as a function of projected distance from the cluster centre. Although the most recent burst will generally be superposed on 
an older population, the Balmer age is strongly weighted toward the youngest stars present.
If the mass-fraction of the most recent burst is very small, the quenching redshifts plotted here are upper limits.
The assumed cosmology has parameters $(\Omega_M,\Omega_\Lambda,h)=(0.3,0.7,0.7)$. 
}\label{fig:radquench}
\end{figure}

Figure~\ref{fig:radtrends_ez} presents the equivalent relations for the derived age (estimated from each Balmer index, and
the average age), the metallicity Fe/H and the four light-element abundance ratios (Mg/Fe, C/Fe, N/Fe, Ca/Fe). 
We find a strong (5$\sigma$) correlation towards younger ages at larger radii, 
whether measured with \Hbeta, \HgF\ or \HdF. The size of the effect is 
consistent between age indicators. Using the average of the three ages obtained for each galaxy,
the age--radius correlation is significant at the 6.5$\sigma$ level, and its slope 
implies a factor of two change in age, per degree in radius. 
The median age for the galaxies beyond 0.4\,deg from the centre
is 3.8\,Gyr, corresponding to their star-formation being quenched at $z\la0.3$ (Figure~\ref{fig:radquench}).

The iron abundance, Fe/H shows a significant (2.7$\sigma$) correlation $0.20\pm0.07$\,dex\,deg$^{-1}$, 
with higher metallicity at larger radius. (Note that the absence of a trend in Fe5270, for example, reflects 
the compensating effect of the age and Fe/H gradients.)
The Mg/Fe abundance ratio is significantly smaller at large radius by
$0.13\pm0.04$ dex\,deg$^{-1}$. Note that the magnesium abundance [Mg/H] = [Mg/Fe] + [Fe/H] 
is consistent with being independent of location in the cluster; thus the variation in [Mg/Fe] is
driven by the greater iron abundance in the outer galaxies, not by lower magnesium. 
The other abundance ratios are also smaller at larger radius, 
but with lower significance levels, $\sim$2$\sigma$. 
The abundances N/H and Ca/H are, like Mg/H, consistent with no radial variation, while C/H may be 
marginally increasing with radius (2$\sigma$). 

Magnesium is produced mainly by Type II supernovae, while iron is released primarily by Type Ia supernovae, 
which are delayed with respect to star formation. Therefore the smaller Mg/Fe (and larger Fe/H) at large radius 
suggest the outer galaxies had extended star-formation histories prior to their quenching. 
In this scenario, it is qualitatively expected that a trend to younger ages would be 
accompanied by a trend towards lower Mg/Fe. Quantitatively however, 
galactic chemical evolution models are required to determine
what star-formation and enrichment histories could reproduce the observed behaviour. 

Finally, we note there is a significant radial colour correlation for the sample, with a slope
$-0.05\pm0.01$\,mag\,deg$^{-1}$ in $g-r$ (bluer colour at larger radius). A small shift, with the same sense,
was previously noted by Terlevich, Caldwell \& Bower (2001) and Carter et al. (2002). 
The galaxies beyond 0.4\,deg have higher recession velocities by 340$\pm$220\,\kms\ than those in the core. 
Preliminary profile fits to SDSS images do not reveal significant radial trends in morphology.

\section{Discussion and summary}

Several authors have claimed to observe significant variation in absorption line strengths
as a function of radius in clusters, with much of this work focused on the Coma Cluster. 

Caldwell et al. (1993) were the first to note a large fraction of galaxies with strong Balmer absorption
in the south-west part of Coma. Their sample probed a smaller radial extent (out to $\sim$50\,arcmin) than our work, 
and was restricted to brighter galaxies ($r\la16$). Their result has driven other work in the south-west region, influencing the
field placement of subsequent studies including the current one. 

\begin{figure}
\vskip 3mm
\includegraphics[angle=270,width=85mm]{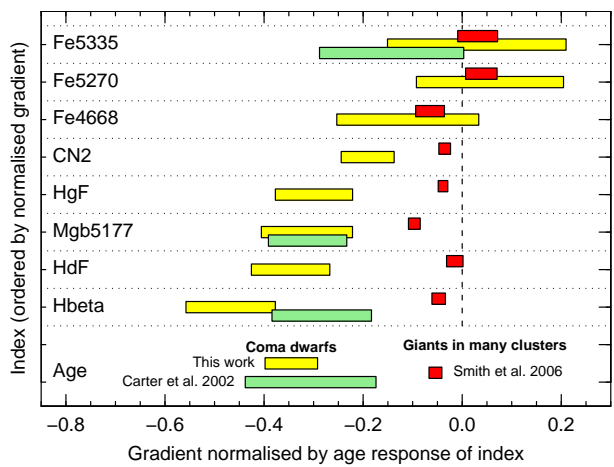}
\vskip -0mm
\caption{Comparison of our index gradient pattern (yellow) with previous work. For clarity, the results for each index have been normalised by its age response. 
Thus if age was the only variable changing with radius, all indices should give the same gradient. Yellow boxes show the $1\sigma$ range from
this work. For Carter et al. (2002, green) we compare their $\langle$Fe$\rangle$ with our Fe5335; 
their Mg2 gradient has been converted to an equivalent gradient in Mgb5177. For Smith et al. (2006, red), we have converted their gradients in 
$R_{200}$ to degrees assuming $R_{200}\approx80$\,arcmin for Coma.
The bottom row shows the age gradients derived for the three studies, including a new estimate for Carter et al., based on their published index gradients. 
}\label{fig:gradboxes}
\end{figure}

Carter et al. (2002) observed Coma galaxies over a wide luminosity range, $r\sim13-19$, in the cluster core and an outer field in
the south-west. The fainter part of their sample corresponds to typical luminosities in our MMT work, and the radial extent is comparable. 
They found a significant radial decrease in the Mg$_2$ index 
at fixed luminosity, interpreting the
result as a metallicity gradient, with lower metallicities in the cluster outskirts. However, their 
data for iron-dominated indices do not share the radial trend, while \Hbeta\ is found to increase with radius. 
Taken together in fact, the results favour an interpretation in terms of age, rather than metallicity. 
Figure~\ref{fig:gradboxes} demonstrates good agreement between Carter et al. (2002) and the present work, 
at the level of the index gradients.  
Our independent data, of much higher signal-to-noise, strongly confirm the earlier results for dwarfs in the south-west part of Coma. 

For giant galaxies, outside of Coma, the situation appears quite different. 
Smith et al. (2006) analysed line-strengths for $\sim$3000 galaxies in $\sim$90 clusters of the 
NOAO Fundamental Plane Survey (NFPS). The NFPS is dominated by $M^\star$ galaxies, i.e. with luminosities 15--40 times larger
than our MMT sample. Smith et al. found stronger Balmer lines and weaker light-element features (Mgb5177, CN2, Fe4668) 
at larger distance from the cluster centres. The iron-dominated indices show no radial dependence.
This pattern can be reproduced by variations in both the age and the $\alpha$-element abundance
ratio \afe\ (comparable to [Mg/Fe]), with no change in the overall metallicity \zh\ (comparable to [Mg/H] here).

The NFPS index gradients are much shallower than those found here or by Carter et al. (Figure~\ref{fig:gradboxes}). 
The relative pattern of gradients is also different, with stronger gradients in the light-element indices than in the 
Balmer lines. This is reflected in the gradients of derived parameters: the Hectospec age gradient (0.35\,dex\,deg$^{-1}$) is seven times larger than 
than the NFPS trend (0.05\,dex\,deg$^{-1}$), while the Mg/Fe trend (0.12\,dex\,deg$^{-1}$) 
is only a factor of 2.4 steeper than the NFPS trend in \afe\ (0.04\,dex\,deg$^{-1}$).

In conclusion, we have confirmed earlier indications that the spectra of faint red-sequence galaxies in Coma are correlated strongly with
their distance from the cluster centre, at least toward the south-west.  
Our analysis, using the latest stellar population models, provides a clear interpretation of the observed trends as being primarily due to age. 
Since such strong radial trends are not seen in general for giant galaxies in
clusters, a key question is whether the difference arises only from the different luminosity ranges, or whether the south-west region of 
Coma is unrepresentative. If the latter, we may be observing the descendents of star-bursting galaxies in cluster-feeding filaments at $z\sim0.2-0.3$
(e.g. Fadda et al. 2008). 
Future work must test whether dwarfs in other clusters, or in other parts of the Coma cluster exhibit similarly strong cluster-centric correlations. 
As one hint that the effect may not be confined to this region, Michielsen et al. (2008) have observed a sample of 18 dwarf ellipticals in Virgo,
and find only young ($\sim$2\,Gyr) objects beyond a radius of $\sim$1\,Mpc, equivalent to $\sim$0.6\,deg at Coma. 
A definitive answer for Coma itself should result from an extension to our Hectospec programme, which will obtain data in 
more fields to provide a complete azimuthal coverage to a radius of 1.5\,deg, or $\sim$2.5\,Mpc.

\section*{Acknowledgments}

RJS was supported by PPARC/STFC rolling grant PP/C501568/1. 
We thank Jenny Graves and Ricardo Schiavon for advice in using \eza. We are grateful to Richard Cool for
making {\sc hsred} publicly available (http://mizar.as.arizona.edu/rcool/hsred), and for advice on Hectospec data reduction. 
{}

\end{document}